\documentclass[aps,prb,twocolumn]{revtex4-1}
\usepackage{dcolumn,amsfonts,amsmath,amssymb}
\usepackage{bm}

\usepackage{graphicx}
\begin{document}
\title{Dynamics of dissipative multiple exciton generation in nanocrystals}
\author{Maryam Azizi}
\author{Pawe{\l} Machnikowski}
\affiliation{Institute of Physics, Wroc{\l}aw University of Technology, 50-370 Wroc{\l}aw, Poland}
\date{\today}

\begin{abstract}
\label{abstract}

The population dynamics of single- and bi-exciton states in semiconductor 
nanocrystals is modeled numerically in the presence of Coulomb coupling 
between single- and two-exciton states and a dissipation channel in order 
to study the transient bi-exciton population that occurs in an optically 
excited semiconductor nanocrystal. The results show that the system evolution 
strongly changes if the dissipation is included. In a certain range of parameters, 
the growth of the exciton number (MEG process) is fast (on picosecond time scale) 
and the following decay (Auger process) is much slower (hundreds of picoseconds). 
In some cases, the maximum occupation of the bi-exciton state increases when 
dissipation is included. The dynamics of an ensemble of nanocrystals with a 
certain size dispersion is studied by averaging over the energy of the bi-exciton 
state which can be different for each single nanocrystal. The validity of Markov 
and secular approximation is also verified.

\end{abstract}

\maketitle

\section{Introduction}
\label{introduction}
One of the possible ways of improving the efficiency of the existing
solar cells is to exploit the process of multiple exciton generation
(MEG) in semiconductor nanocrystals (NCs) \cite{nozik02}.  Such an 
effect consists in generation of two or more electron-hole pairs by a
single high energy photon and thus converts the excess above-bandgap 
energy into useful current. This process is enabled by Coulomb
coupling between single-pair (exciton, X) states and two-pair
(biexciton, BX)
states in a NC (or, in general, between states with $n$ and
$n+1$ pairs) and consists in an intraband relaxation of a carrier
(typically an electron, due to larger energy scales of confined states
in the conduction band) accompanied by a creation of a new
electron-hole pair (an inverse Auger process). In this way, the excess
energy obtained by an electron upon absorbing a high-energy photon is
not dissipated in a phonon 
relaxation processes and becomes available for photovoltaic conversion.

The initial experimental results, showing very high values of
the quantum efficiency of photoconversion in various systems
\cite{schaller04,schaller05,ellingson05,schaller06,schaller07,pijpers07,beard07,ji09},
were subsequently reinterpreted
\cite{nair07,trinh08,pijpers08,ben-lulu08,nair08} based on the growing
understanding of 
the experimental difficulties that might lead to overestimating the
achieved numbers of excitons per single absorbed photon
\cite{mcguire08,mcguire10,binks11}.
Nonetheless, more recent experiments on real 
NC-based solar cell devices \cite{sambur10,semonin11} do provide a
direct proof of  
the usefulness of this process in solar energy conversion.
Theoretically, the description of the X and BX spectrum and the X-BX
couplings that are essential for the MEG process has been proposed
using the methods of density functional \cite{deuk11,deuk12},
pseudopotential 
\cite{franceschetti06,rabani08,califano09,baer12}, 
tight binding \cite{allan06,delerue10,korkusinski10,korkusinski11},
and k$\cdot$p theory 
\cite{shabaev06,witzel10,silvestri10}. 

Along with the investigation of
these structural properties, much attention has been devoted to the
carrier dynamics in a nanocrystal under optical excitation at
energies high enough to generate multiple excitons, in
particular to the role of decoherence and relaxation. These studies
included dynamical simulations of few-level models
\cite{shabaev06,schulze11} as well as of many-level models aiming at
reproducing the density of X and BX states in the
high-energy sector of a nanocrystal 
\cite{witzel10,rabani10,piryatinski10,velizhanin11a,korkusinski11}. In
many cases,
dissipative effects are included in these models on a phenomenological
level and expressed by a number (usually small) of dephasing rates
\cite{shabaev06,allan06,witzel10,schulze11}. In 
this way, the multiple exciton generation could be described as a
process competing with exciton relaxation \cite{shabaev06,witzel10}
and, in certain cases, suppressed by coupling to the dissipative
environment (typically considered to be phonons) \cite{allan06}. 

In this paper, we study the time evolution of the X-BX
system within a minimal, three-level model that accounts both for the
impact ionization and Auger recombination in the presence of
dissipation. Expressing the couplings to the environment in a generic
form in terms of a physically motivated set of spectral densities
allows us to characterize the 
emerging coupling to the Coulomb-correlated X-BX
eigenstates and to discuss the dependence of the rates of various
phonon-assisted processes (relaxation and impact ionization) on the
Coulomb coupling itself. We show that, on the general level, the dissipative
MEG process is determined by the same couplings to the dissipative environment as the
carrier relaxation and dephasing.
Furthermore, we find out that the system dynamics realizes
various dynamical scenarios, depending on the alignment of the X
and BX levels and on the relation between the level spacing and
the spectral properties of the coupling to the environment (in
particular, the high-frequency cut-off of the spectral density). As we
show, the presence of dissipation considerably modifies the system
dynamics and, in many cases, increases the efficiency of the MEG
process. We study also the role of the excitation conditions (pulsed,
continuous wave or incoherent thermal) and show that the strong differences
between the system kinetics under different excitation conditions
\cite{schulze11} are washed out by dissipation. Finally, we assess the
validity of Markov and secular approximations for the description of
dissipation-assisted MEG in nanocrystals.

The paper is organized as follows. In Sec.~\ref{sec:model}, we define
the model (Sec.~\ref{sec:system}), describe the Master equations for the
system evolution (Sec.~\ref{sec:evol}), and discuss the formal structure
of the carrier-environment coupling in a NC (Sec.~\ref{sec:env}). In
Sec.~\ref{sec:results}, the results of our simulations are discussed:
first the dynamics in the Markov limit is studied (Sec.~\ref{sec:Markov}) and then
non-Markovian corrections are discussed
(Sec.~\ref{sec:non-Markov}). Finally, Sec.~\ref{sec:concl} concludes
the paper.

\section{Model}
\label{sec:model}

\subsection{The system}
\label{sec:system}
\begin{figure}[tb]
\centering
\includegraphics[width=85mm]{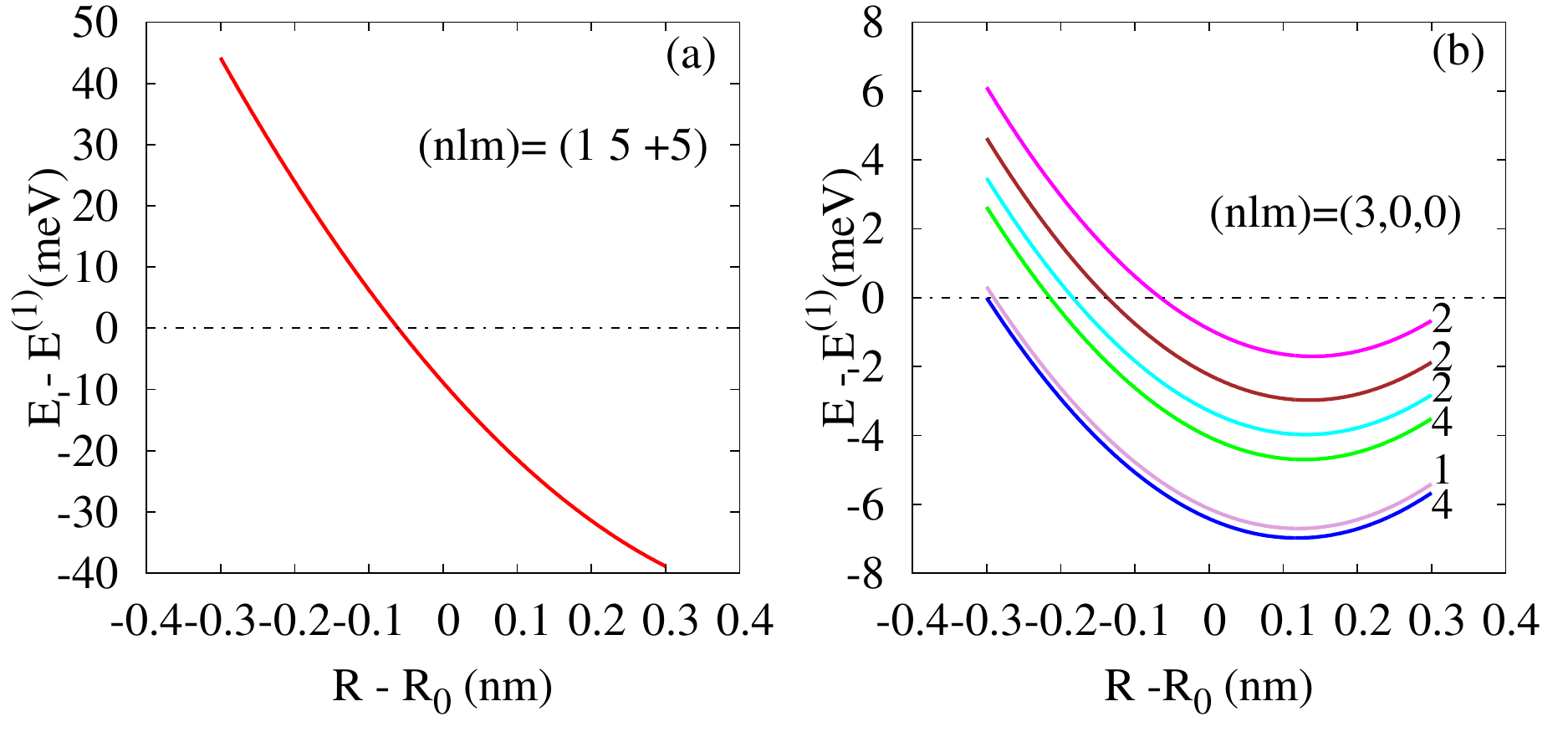}
\caption{\label{Fig0_MA}(Color online)
Relative energy of BX states in vicinity of selected bright X states 
with quantum numbers as indicated as a function of the NC radius 
around $R_0=3~$nm. The numbers on the right denote degeneracy.} 
\end{figure}
Although the density of BX states in a nanocrystal is very high, only 
a few tens of them are coupled to a given bright X state \cite{kowalski09}. 
Moreover, usually only a few coupled BX states lie in the vicinity of the X 
state while vast majority of them is relatively distant. In Fig.~\ref{Fig0_MA}, 
we show two selected examples of the spectral positions of BX states relative 
to selected X states to which they are coupled (the quantum numbers for the 
selected exciton states are the same for the electron and the hole and are 
shown in the figure). This approximate result is obtained within the single-band 
envelope function approximation with Coulomb interactions included in the lowest 
order for an InAs nanocrystal with the radius R close to $R_0=3~$nm~\cite{kowalski09,brus84}. 
As can be seen, only one or a few BX states appear in the spectral vicinity of 
a given X state. Thus, the essential features of the MEG dynamics can be 
expected to be determined by impact ionization within groups of a few states. 

\begin{figure}[tb]
\centering
\includegraphics[width=85mm]{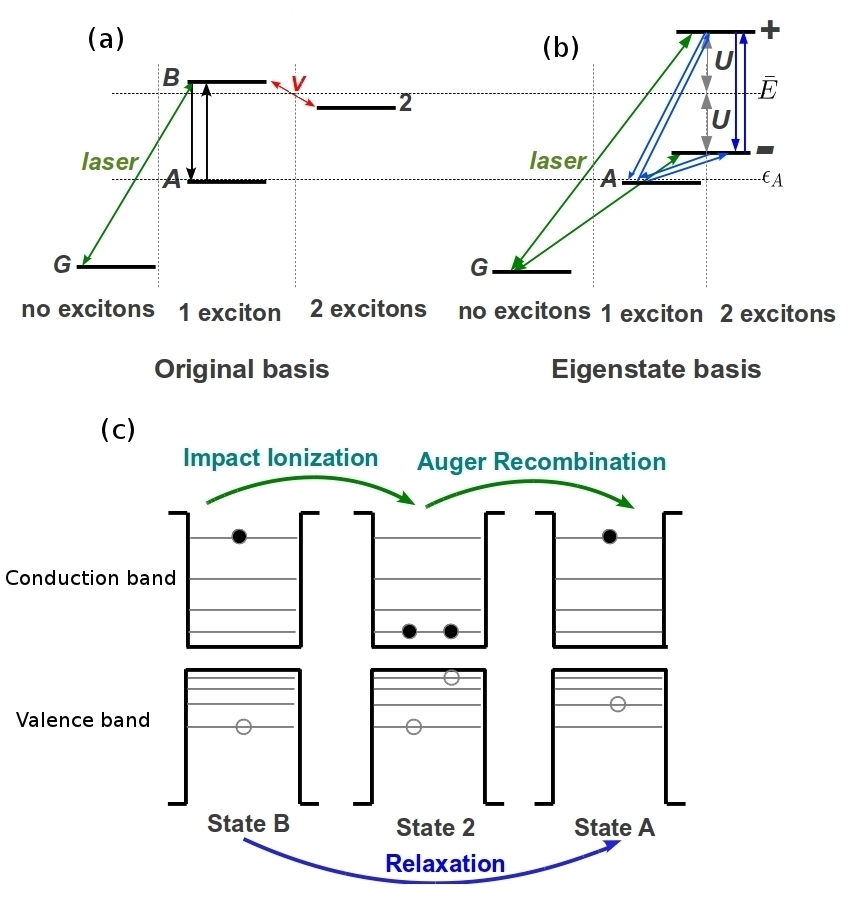}
\caption{\label{Fig1_MA}(Color online)
(a) The schematic diagram of system states and couplings. The green 
arrow shows the lase excitation while the black arrows represent the 
environment-induced transitions. The Coulomb coupling between X and 
BX configurations is shown with a red arrow. (b) The diagram of 
system eigenstates with the optical excitation paths (green) and 
environment-induced transitions (blue). (c) Graphical representation 
of the processes involved.} 
\end{figure}

Consider an optically active (bright) excited X state of a NC. At lower
energies, other X states are present to which the carriers can
relax. For our purpose, it is sufficient to consider one such
level. We will consider situations in which a BX state Coulomb-coupled
to the bright X state is present in between the two X states, which is a
common situation for highly excited X states, where the coupled BX
states are rather dense. Taking into account the selection rules for
interband Coulomb couplings\cite{kowalski09}, it is rather unlikely
that this BX state will also be coupled to the other, lower X
state. Therefore, for our dynamical modeling, we consider a four-level model 
of a nanocrystal, as shown in the Fig.~\ref{Fig1_MA}(a). 
Here the state G refers to the ground state (empty nanocrystal), A and B 
denote the two X states and 2 is the BX state. We assume that the BX state 
2 is coupled only to the state B by a Coulomb coupling $V$ which is taken to be a real parameter. 

The Hamiltonian of the carrier system is then
\begin{eqnarray}
H_0=\epsilon_A|A\rangle\! \langle A|+\epsilon_B|B\rangle\! \langle B| +\epsilon_2 |2\rangle\! \langle 2| \nonumber\\+V(|B\rangle\!\langle2| +|2\rangle\! \langle B|),
\label{laser}
\end{eqnarray}
where $\epsilon_A$,$\epsilon_B$ and $\epsilon_2$ are the energies 
of the states A, B and 2 respectively. We set the energy of the ground state to be zero.
Moreover, the system is excited by a classical light pulse which 
has the frequency $\Omega$ close to the $G\longleftrightarrow B$ 
transition. By standard selection rules, this pulse couples the 
ground state only to the X states. We assume that only the state B 
is bright. The corresponding Hamiltonian of the excitation is
\begin{eqnarray}
H_{\mathrm{las}}=\frac{1}{2}f(t)\left(|G\rangle\!\langle B| e^{i\Omega t} +|B\rangle\!\langle G| e^{-i\Omega t}\right).
%\label{laser}
\end{eqnarray}
In addition, the system undergoes dissipative dynamics due to the 
interaction with its environment. We do not assume any specific 
form of the coupling to the environment and aim at a model which 
is independent of the nature of this coupling. The only restriction 
on this coupling is that in the absence of Coulomb-induced mixing 
between the X and BX configurations no interband processes are allowed 
[see Fig.~\ref{Fig1_MA}(a)]. Thus, the interaction with the environment 
can be described by the following Hamiltonian
\begin{eqnarray}\label{Hint1}
H_{\mathrm{int}}=|A\rangle\! \langle A|\mathcal{R}_{AA}+|B\rangle\! \langle B|\mathcal{R}_{BB}+|2\rangle\! \langle 2|\mathcal{R}_{22} \nonumber\\+|A\rangle\! \langle B|\mathcal{R}_{AB}+ |B\rangle\! \langle A|\mathcal{R}_{BA},
%\label{laser}
\end{eqnarray}
where $\mathcal{R}_{\alpha\beta}$ (with $\alpha, \beta=A,B,2$) are 
certain environment operators with the property $\mathcal{R}_{\alpha\beta}=\mathcal{R}_{\beta\alpha}^{\dag}$.

The system evolution is described in the basis of eigenstates 
of $H_0$: $|G\rangle$, $|A\rangle$, $|+\rangle$ and $|-\rangle$, 
where $|+\rangle$ and $|-\rangle$ result from the Coulumb coupling 
between the X state $|B\rangle$ and the BX state $|2\rangle$, 
\begin{eqnarray}
|+\rangle =\ \ \cos(\theta/2) |B\rangle+\sin(\theta/2) |2\rangle,\nonumber \\  
|-\rangle =-\sin(\theta/2) |B\rangle+\cos(\theta/2) |2\rangle,
\label{pm}
\end{eqnarray}
Here, $\theta$ is the mixing angle, defined by $\tan\theta=2V/(\epsilon_B - \epsilon_2)$, 
which is close to $0$ for weakly mixed X and BX states and goes to $\pm\pi/2$ 
if a nearly degenerate pair of X and BX states is strongly coupled.
The energies of these states are $E_\pm=\bar{E}\pm U$, where 
$\bar{E}=\left(\epsilon_B + \epsilon_2\right)/2$ and $U=[\left(\epsilon_B - \epsilon_2\right)^2/4+V^2]^{1/2}$.
In the eigenstate basis, the interaction Hamiltonian can be written as
\begin{eqnarray}\label{Hint}
H_{\mathrm{int}}=\sum_{i,j=A,\pm}\sigma_{ij}\widetilde{\mathcal{R}}_{ij},
\end{eqnarray}
where $\sigma_{ij}=|i\rangle\! \langle j|$ and

\begin{subequations}\begin{align} \label{R1}
\widetilde{\mathcal{R}}_{A+}&=\cos\frac{\theta}{2}\ \mathcal{R}_{AB},\quad \widetilde{\mathcal{R}}_{A-}=-\sin\frac{\theta}{2}\ \mathcal{R}_{AB},\\
\widetilde{\mathcal{R}}_{\pm\pm}&=\frac{1}{2}\left(\mathcal{R}_{BB}+\mathcal{R}_{AA}\right) \pm \frac{1}{2}\cos\theta\left(\mathcal{R}_{BB}-\mathcal{R}_{AA}\right), \\
\widetilde{\mathcal{R}}_{+-}&=\frac{1}{2}\sin\theta\left(\mathcal{R}_{22}-\mathcal{R}_{BB}\right), 
\end{align}\end{subequations}

with $\widetilde{\mathcal{R}}_{ij}^+=\widetilde{\mathcal{R}}_{ji}.$

\subsection{Evolution: Master equation}
\label{sec:evol}
While Markov approximation has commonly been used to model the
dissipation effect on the MEG process in nanocrystals (often on the level of
phenomenological dephasing rates)
\cite{shabaev06,schulze11,witzel10,rabani10,piryatinski10,velizhanin11a,korkusinski11},
its validity is not obvious for the present system. Indeed, as
observed in experiment and reproduced by our simulations discussed
below, only the Auger recombination phase of the system dynamics is slow,
while the initial impact ionization dynamics takes place on much
shorter, picosecond time scales. Therefore, in this paper, we will
compare the Markovian dynamics with a more general approach, where the
reservoir memory is taken into account. In both cases, the evolution
of the system will be described in the density matrix
formalism by solving the appropriate quantum Master equation. 

Thus, the starting point for our modeling of the system evolution is the
time-convolutionless Master equation for the
reduced density matrix of the charge subsystem in the NC 
in the lowest order. In the 
interaction picture, this equation has the form \cite{breuer02}
\begin{eqnarray}
\dot{\tilde{\rho}}=-\frac{1}{\hbar^{2}} \mathrm{Tr}_{\mathrm{R}}\int_{0}^{t} d\tau 
\left[ H_{\mathrm{int}}(t),\left[
    H_{\mathrm{int}}(\tau),\tilde{\rho}(t)\otimes\rho_{\mathrm{R}} 
\right]  \right],
\end{eqnarray}

where $H_{\mathrm{int}}(t)$ and $\tilde{\rho}$ are 
the interaction Hamiltonian and the reduced density matrix of the nanocrystal in 
the interaction picture, $\rho_{\mathrm{R}}$ is 
the density matrix of the reservoir at
thermal equilibrium, and $\mathrm{Tr}_{\mathrm{R}}$ denotes the partial trace
over the reservoir degrees of freedom.
Upon substituting the interaction Hamiltonian from Eq.~\eqref{Hint} and
taking the partial trace
%and transforming to the Schr\"odinger picture, 
this yields
\begin{eqnarray}\label{partialform}
\dot{\tilde{\rho}}(t) & = & -\frac{1}{2} 
e^{i(\omega_{i}-\omega_{j}+\omega_{k}-\omega_{l})t} \nonumber\\
&& \times \Gamma_{ijkl}(t)\left[ 
\sigma_{ij}\sigma_{kl}\tilde{\rho} (t)
-\sigma_{kl}\tilde{\rho} (t)\sigma_{ij}  \right]
+\mathrm{h.c.},
\end{eqnarray}
with $\omega_i=E_i/\hbar$ and the time-dependent rates
\begin{equation}
\Gamma_{ijkl}(t)=\frac{2}{\hbar^2}\mathrm{Re}\sum_{ijkl} \int_{0}^{t} ds e^{i(\omega_{l}-\omega_{k})s}\left\langle \widetilde{\mathcal{R}}_{ij}(s) \widetilde{\mathcal{R}}_{kl}\right\rangle,
\label{rates-t}
\end{equation}
where 
$\widetilde{\mathcal{R}}_{ij}(s)$ denotes the operator
$\widetilde{\mathcal{R}}_{ij}$ in the interaction picture with respect
to the reservoir Hamiltonian and we have neglected the imaginary parts
of the rates that describe reservoir-induced energy shifts. 

The reservoir correlation function (``memory function'')
$\langle \widetilde{\mathcal{R}}_{ij}(s) \widetilde{\mathcal{R}}_{kl}
\rangle$ is related to the spectral density
\begin{eqnarray}\label{spectral}
\widetilde{R}_{ijkl}(\omega)=\frac{1}{2\pi\hbar^2}\int dt e^{i\omega t} \langle{\widetilde{\mathcal{R}}}_{ij}(t){\widetilde{\mathcal{R}}}_{kl}\rangle,
\end{eqnarray}
for $i,j,k,l=A,\pm$, which fully characterizes the dissipative coupling to the
environment. In the same way, spectral densities 
$\mathcal{R}_{\alpha\beta\gamma\delta}(\omega)$ are defined in terms
of correlation functions between the operators $R_{\alpha\beta}$ in
the original basis [Eq.~\eqref{Hint1}]. 
If the reservoir correlations decay on a
certain time scale (reservoir memory time) then, on longer time
scales, the rates become constant and equal to
\begin{displaymath}
\Gamma_{ijkl}(t)\stackrel{t\to\infty}{\longrightarrow}\Gamma_{ijkl}(\infty)=
2\pi R_{ijkl}\left(\omega_{l}-\omega_{k}\right).
\end{displaymath}
Moreover, as follows from Eq.~(\ref{rates-t}), in the absence of
degeneracy, the rates other than
$\Gamma_{ijji}$ and $\Gamma_{iijj}$ contain an oscillating factor and 
can be assumed to be small if the separation of the levels is
large and the overall system dynamics is slow. Therefore, it is 
common to use the secular approximation where the terms containing
such oscillating rates are neglected. The rates of the type
$\Gamma_{iijj}$ at long times are proportional to the corresponding
spectral density 
at zero frequency, which vanishes in many common situations
(super-Ohmic reservoirs and Ohmic reservoirs at zero temperature, see
below). Thus, one reaches the
commonly used Markov approximation with the evolution equation in
the Lindblad form \cite{breuer02}, which in the Schr\"odinger picture
can be written as
\begin{eqnarray}\label{Lindblad}
\dot{\rho}=-\frac{i}{\hbar}[H_0,\rho]+\sum\limits_{ij}\Gamma_{ij}(\sigma_{ji}\rho\sigma_{ij}-\frac{1}{2}\{\sigma_{ij}\sigma_{ji},\rho\}),
\end{eqnarray}
where $\Gamma_{ij}=\Gamma_{ijji}(\infty)$.

\subsection{Environment: Spectral densities}
\label{sec:env}

Obviously, the spectral densities defining the transition rates are related 
to those in the original basis. In particular, 

\begin{eqnarray}\label{R}
\widetilde{R}_{-++-}(\omega)&=&\frac{1}{4}\sin^2\theta[R_{2222}(\omega)\nonumber\\
&&-R_{22BB}(\omega)-R_{BB22}(\omega)+R_{BBBB}(\omega)], \nonumber\\
\widetilde{R}_{A\pm \pm A}(\omega)&=&\frac{1}{2}(1\pm \cos\theta)\ R_{ABBA}, 
\end{eqnarray}
with $\mathcal{R}_{jiij}(\omega)=\mathcal{R}_{ijji}(\omega).$

It is clear that the relevant spectral densities $\widetilde{R}_{ijji}$ 
describing the transitions between Coulomb-correlated X-BX configurations 
are combinations of the spectral densities $R_{\alpha\beta\gamma\delta}$ 
that describe dephasing and interaband relaxation between X and BX states. 
Interestingly, the transition between the two Coulomb-mixed states,
that is, the impact ionization process, described by
$\widetilde{R}_{\mp\pm\pm\mp}$ is entirely related to the diagonal
couplings between the original states and the environment. Obviously,
the corresponding rate vanishes for small X-BX mixing as
$\theta^{2}\sim [V/(\epsilon_{B}-\epsilon_{2})]^{2}$. On the other
hand, the transition to the other X state A (relaxation or impact ionization) is governed by the
off-diagonal couplings, which are related to intraband relaxation
between these states. If the diagonal and off-diagonal couplings are
of similar magnitude (as it is the case, e.g., for carrier-phonon
couplings), then the impact ionization and Auger recombination are
formally suppressed by a factor $\sin^{2}\theta$ or
$\sin^{2}\theta/2$ as compared to relaxation. However, as we will show
below, what really matters is the frequency dependence of the
reservoir spectral density which can make the impact ionization
process favorable, depending on the X-BX level alignment.  

The details of the system-environment coupling may vary for different 
nanocrystal systems and, to our knowledge no general microscopic theory 
has been proposed for its exact treatment. Hence, in most of our simulation, we take the spectral 
densities in the original basis in the simplest Ohmic form, $R_{\alpha\beta\gamma\delta}(\omega)=a_{\alpha\beta\gamma\delta}\left[n_B(\omega)+1\right]J(\omega)$, where $J(\omega)=(J_0 \omega/\omega_0) \exp[-(\omega/\omega_0)^2]$ (see Sec.~\ref{sec:non-Markov} for comparison with a super-ohmic case).

Here, $n_B(\omega)$ is the Boltzmann distribution function, $J_0$ 
is the overall magnitude of the dissipation and $\omega_0$ is the 
cut-off frequency. In the simplest case of lowest-order acoustic 
phonon processes, $\omega_0$ is on the order of R/c, where c is the 
speed of sound, which yields a value on the order of $\mathrm ps^{-1}$. 
This can be different if multiple-phonon processes are included.
The values of the coefficients $a_{\alpha\beta\gamma\delta}$ follow
from the physical nature of the carrier-environment coupling:
The spectral densities $R_{\alpha\alpha\beta\beta}(\omega)$ result
from diagonal couplings between the carriers in the NC and their
environment. It seems reasonable to assume that these couplings are
at least approximately proportional to the charge density (this is
true, e.g., for carrier-phonon couplings as well as for Coulomb
couplings to fluctuating background or impurity charge). Thus, 
typically, $\mathcal{R}_{22}=2\mathcal{R}_{BB}$ and, in consequence,
$a_{22BB}=a_{BB22}=2, a_{2222}=4$ and $a_{\alpha\beta\gamma\delta}=1$ otherwise. 

As we will show, the system dynamics in the dissipative MEG process 
depends to some extent on the excitation conditions. Nevertheless, 
instead of including the electromagnetic field explicitly in our 
simulation we note that the excitation may fall essentially in three 
classes: a short (spectrally broad) pulse excites 
an optically active (bright) X state, a long (spectrally narrow) pulse excites selectively an 
eigenstate of the system and broad band thermal radiation 
excites an incoherent mixture of system eigenstates, proportionally 
to their overlap with the bright X state (see Appendix for details). 
Thus, as the initial state we take the state $|B\rangle$, corresponding 
to an ultra fast coherent excitation (a broad band laser pulse), a pure $|+\rangle$ 
or $|-\rangle$ state for a narrow band laser pulse or  
a mixture of the eigenstates $|+\rangle$ and $|-\rangle$, corresponding 
to incoherent excitation by thermal radiation.

\section{Results}
\label{sec:results}

\subsection{Dissipative MEG dynamics}
\label{sec:Markov}

Several parameters play a crucial role in the dissipative MEG process: 
the energy differences between the X state A, the excited state B and 
the BX state 2, the Coulomb coupling between the X and BX states, and 
the parameters governing the dissipative relaxation process (the 
magnitude $J_0$ and the frequency cut-off $\omega_0$). In this section, 
we study the dynamics of the dissipative MEG process for various 
energetical alignment of the states, assuming fixed values of the 
dissipation parameters, and compared the results to the case without dissipation.

\begin{figure}[tb]
\centering
\includegraphics[width=60mm]{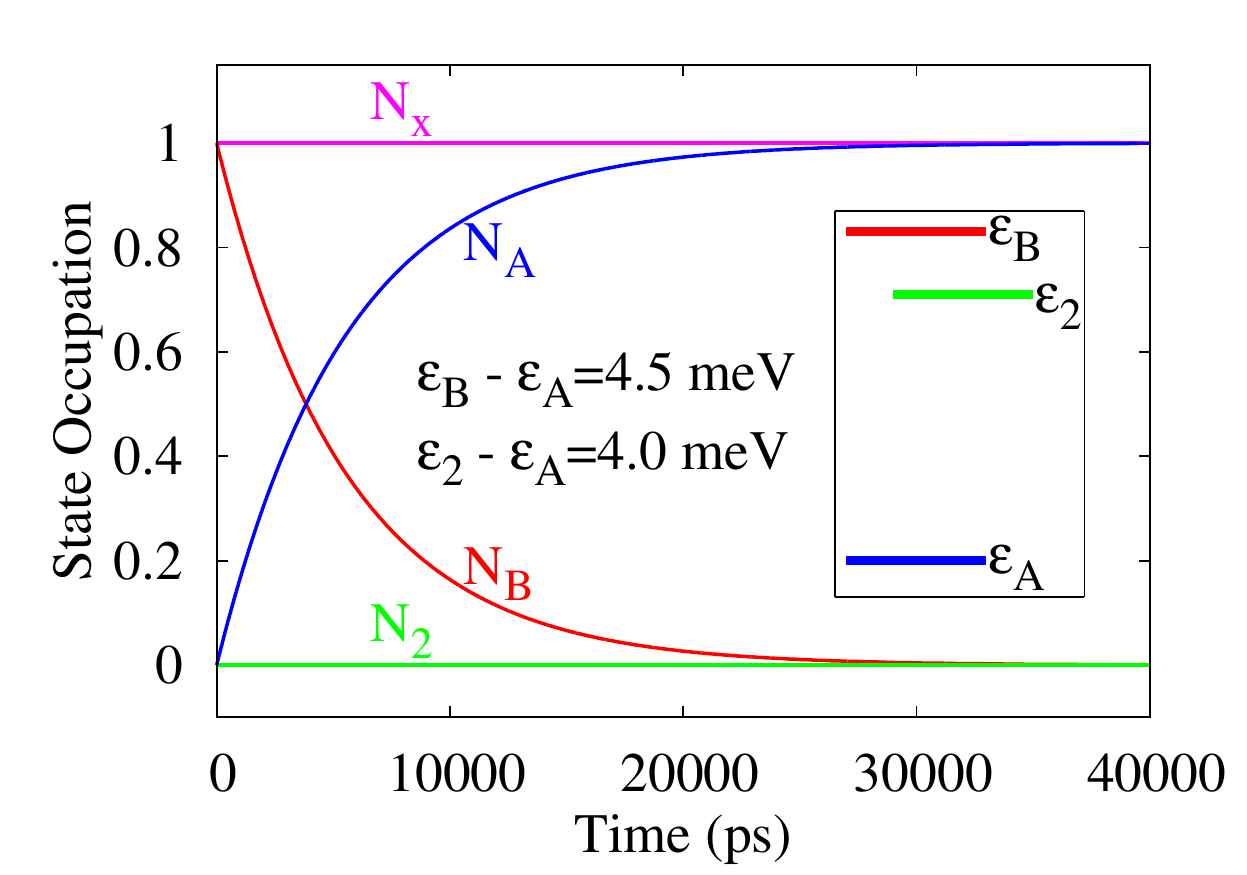}
\caption{\label{Fig2_MA}(Color online)
The variation of the state occupations $N_A$, $N_B$ and $N_2$ 
and the average number of excitons $N_x$ in the absence of Coulomb coupling at $T=0$~K.} 
\end{figure}

In the absence of Coulomb coupling, there is no mechanism 
for a transition to the BX state (this state is completely decoupled), 
hence, only relaxation between the states B and A takes place. 
This is shown in Fig.~\ref{Fig2_MA} (the dynamics in this case 
is independent of the excitation conditions). In this case, the 
BX state is indeed never occupied and the total number of excitons, 
$N_x$, remains equal to one. The only occurring process is the 
intraband occupation transfer from the initial state B to the dark X state A. 

\begin{figure}[tb]
\centering
\includegraphics[width=85mm]{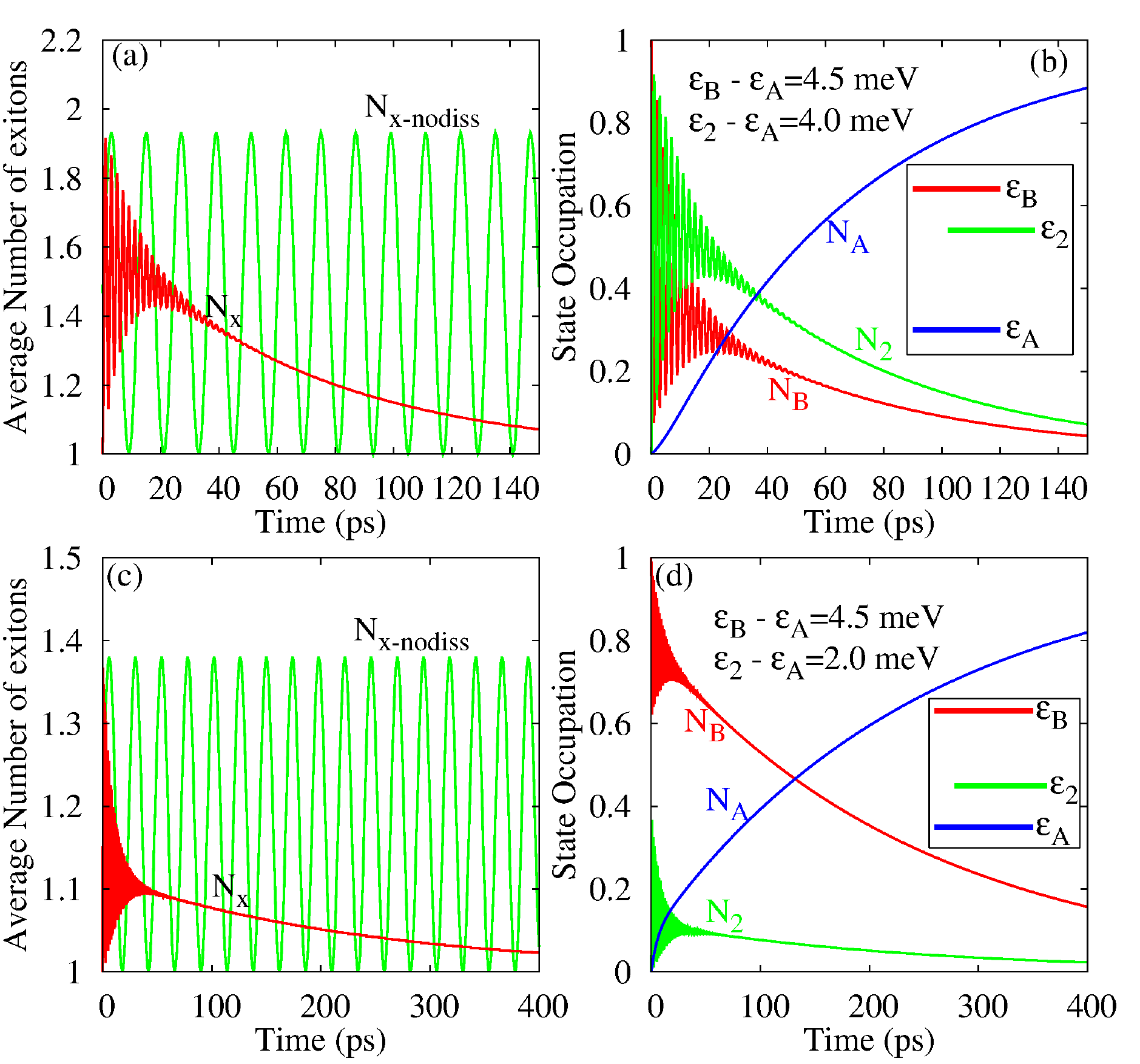}
\caption{\label{Fig3_MA}(Color online)
The occupations of the system states (right panels) 
and the average number of excitons (left panels) for $V_B=1~$meV 
for 2 possible alignments of the energy levels at $T=0$~K. 
In the left panels, the average number of excitons in the case without dissipation. 
Coherent ultrafast excitation is assumed here. } 
\end{figure}

The variation of the state occupations and the average number 
of excitons in the presence of a realistic Coulomb coupling 
($V_B=1$~meV) is shown in Fig.~\ref{Fig3_MA} for the case of 
coherent ultrafast excitation. As mentioned above, under these 
excitation conditions, the initial system state is $|B\rangle$, hence $N_x=1$.
When the bright state $|B\rangle$ and the BX state $|2\rangle$ 
are close to each other, Fig.~\ref{Fig3_MA}(a,b), the Coulomb 
interaction leads to strong mixing of the states $|B\rangle$ 
and $|2\rangle$. As a result, transition between the resulting 
eigenstates are efficient which means that the dissipative impact 
ionization is very fast as manifested by the rapid growth (below 1 ps) 
of the occupation of the state $|2\rangle$ and the corresponding 
increase of the average number of excitons (the lower eigenstate 
is still predominantly bi-excitonic). Since the initial state 
$|B\rangle$ is a superposition of system eigenstates, there is 
an intense oscillation at the beginning of the process but its 
amplitude goes to zero in about 10 ps. Later on the system 
relaxes to the state $|A\rangle$, which corresponds to the Auger 
recombination. The BX state decays in this process on a time scale 
of about 150 ps. The slow rate of the Auger process is due to the 
relatively large energy distance to the state A, which makes the 
relaxation to this state ineffective. 

For another configuration, Fig.~\ref{Fig3_MA}(c,d), when the 
BX state is shifted closer to the state $|A\rangle$, the 
relaxation behavior does not change significantly. However, 
the maximum occupation of the state $|2\rangle$ is much lower. 
This results from the larger energy spacing between the states 
which is beyond the frequency cut-off of the spectral density, 
which considerably suppresses the relaxation to the state $|-\rangle$ 
which is a predominantly BX state. Thus, in the competition between 
the impact ionization (transition to the state $|-\rangle$ or, almost 
equivalently, to the BX state $|2\rangle$) and the usual relaxation 
(transition to $|A\rangle$), the latter starts to dominate. 
In addition, the similar energy spacings $\epsilon_B-\epsilon_2$ 
and $\epsilon_2-\epsilon_A$ makes the Auger recombination from 
the state 2 to the state A much more effective in comparison 
to the impact ionization. In this case, the oscillation amplitude 
decays in a slightly longer time of a few tens of picoseconds. 
In both of these configurations, the impact ionization competes 
strongly also with the Auger recombination, the latter being 
faster in the second configuration. 

In the absence of dissipation, $J_0=0$, for both the above 
mentioned configurations, the average number of excitons 
oscillates constantly (green lines in Fig.~\ref{Fig3_MA}(a,c)). However, as one could expect, 
the amplitude of the oscillation is different for various 
configurations due to different degrees of mixing between 
the eigenstates. Note that for the first alignment (Fig.~\ref{Fig3_MA}(a,b)), 
the number of excitons generated in the presence of dissipation 
is larger than that achieved without dissipation 
(taking the average of oscillation in both cases).

\begin{figure}[tb]
\centering
\includegraphics[width=85mm]{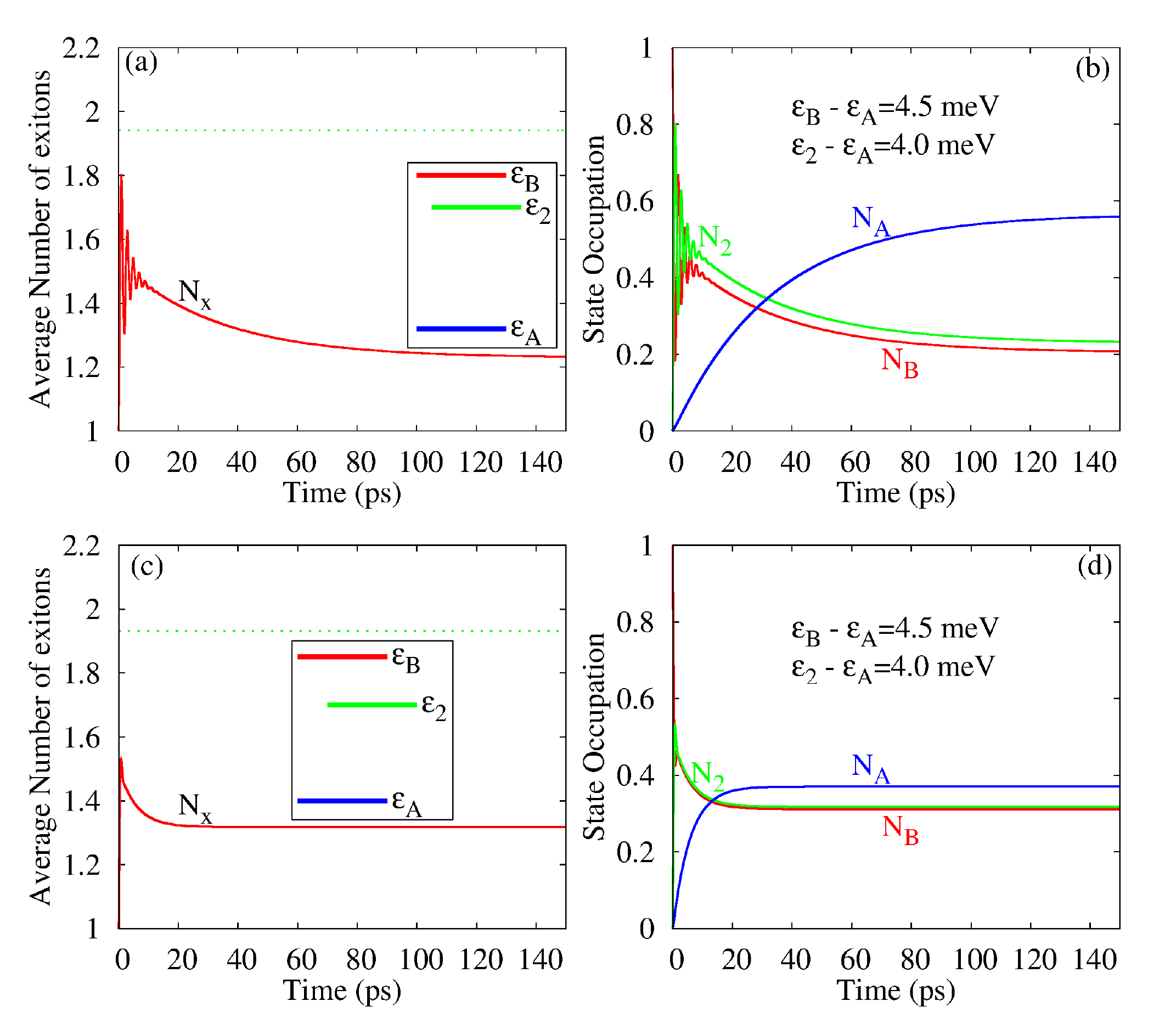}
\caption{\label{Fig350_MA}(Color online)
As in Fig.~\ref{Fig3_MA} but at $T=50$~K (a,b) and $T=300$~K (c,d). } 
\end{figure}

At higher temperature, $T=50$~K, Fig.~\ref{Fig350_MA}(a,b), the relaxation dynamics does not
change considerably but, the oscillation amplitude decays in a shorter time. 
At $T=300$~K, Fig.~\ref{Fig350_MA}(c,d), the initial average number of excitons is slightly decreased and 
the oscillations vanish quickly. The final average number of excitons goes up at higher temperatures 
because of nonzero thermal occupations.

\begin{figure}[tb]
\centering
\includegraphics[width=85mm]{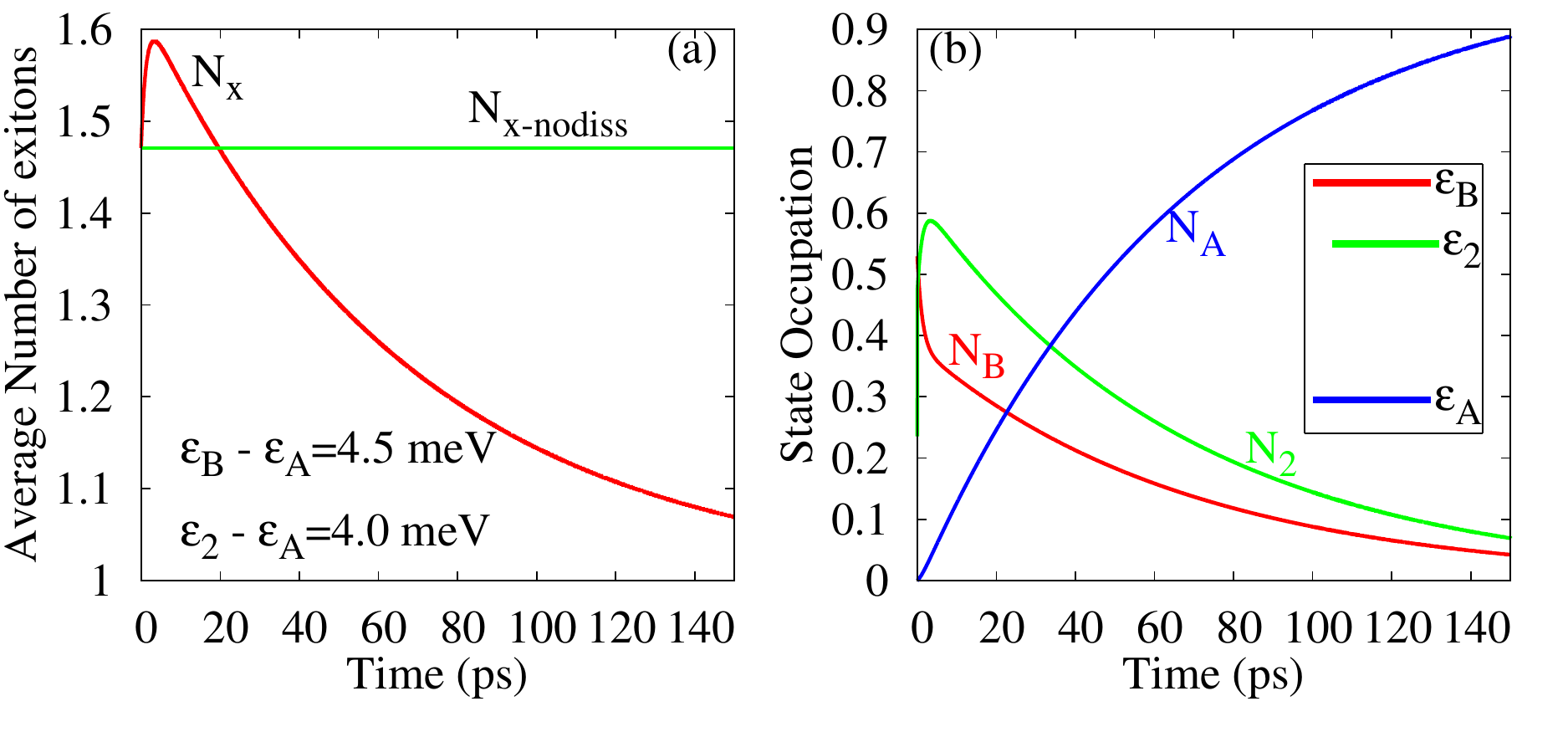}
\caption{\label{Fig4_MA}(Color online)
The dynamics of (a) the average number of excitons (b) 
the system state occupations for incoherent excitation 
by thermal radiation at $T=0$~K. In (a), the average number 
of excitons in the case without dissipation is also 
shown (green line). } 
\end{figure}

Under incoherent excitation, according to the Fermi golden rule, 
the system eigenstates are excited proportionally to their coupling 
to the light, that is, to the admixture of the bright state $|B\rangle$. 
Since in the presence of the Coulomb coupling, the eigenstates 
$|+\rangle$ and $|-\rangle$ are superpositions of the original 
X state $|B\rangle$ and BX state $|2\rangle$, they are characterized 
by the average number of exciton between 1 and 2. Hence, the initial 
number of excitons may exceed 1.
The system dynamics in this case, shown in Fig.~\ref{Fig4_MA} for one 
of the level alignments, is very similar to that following a coherent 
excitation but no oscillations are seen since no coherence between 
the eigenstates is present. There is again a strong competition 
between the impact ionization, the X relaxation and the Auger 
recombination. As a result, the occupation of the BX state again 
grows rapidly and then decays completely on a longer time scale.

Under these excitation conditions, in the absence of dissipation, 
the average number of excitons would remain constant but in the 
presence of dissipation it increases considerably to a maximum 
value and then decays. It is clear that the maximum average number 
of excitons resulting from the dissipative MEG dynamics in this 
case exceeds that following the excitation in the absence of 
dissipation. This is due to the dissipative transition to the 
predominantly biexcitonic state $|-\rangle$, which develop on a 
time scale much shorter than the subsequent Auger transition to 
the lowest state $|A\rangle$.

It can be seen by comparing Fig.~\ref{Fig4_MA} and Fig.~\ref{Fig3_MA} 
that for incoherent excitation, the value of the average number of 
excitons is equal to the mean value of the oscillations that could 
be seen in Fig.~\ref{Fig3_MA}. The same holds true also for 
the other level alignments, not shown in Fig.~\ref{Fig4_MA}.

\begin{figure}[tb]
\centering
\includegraphics[width=85mm]{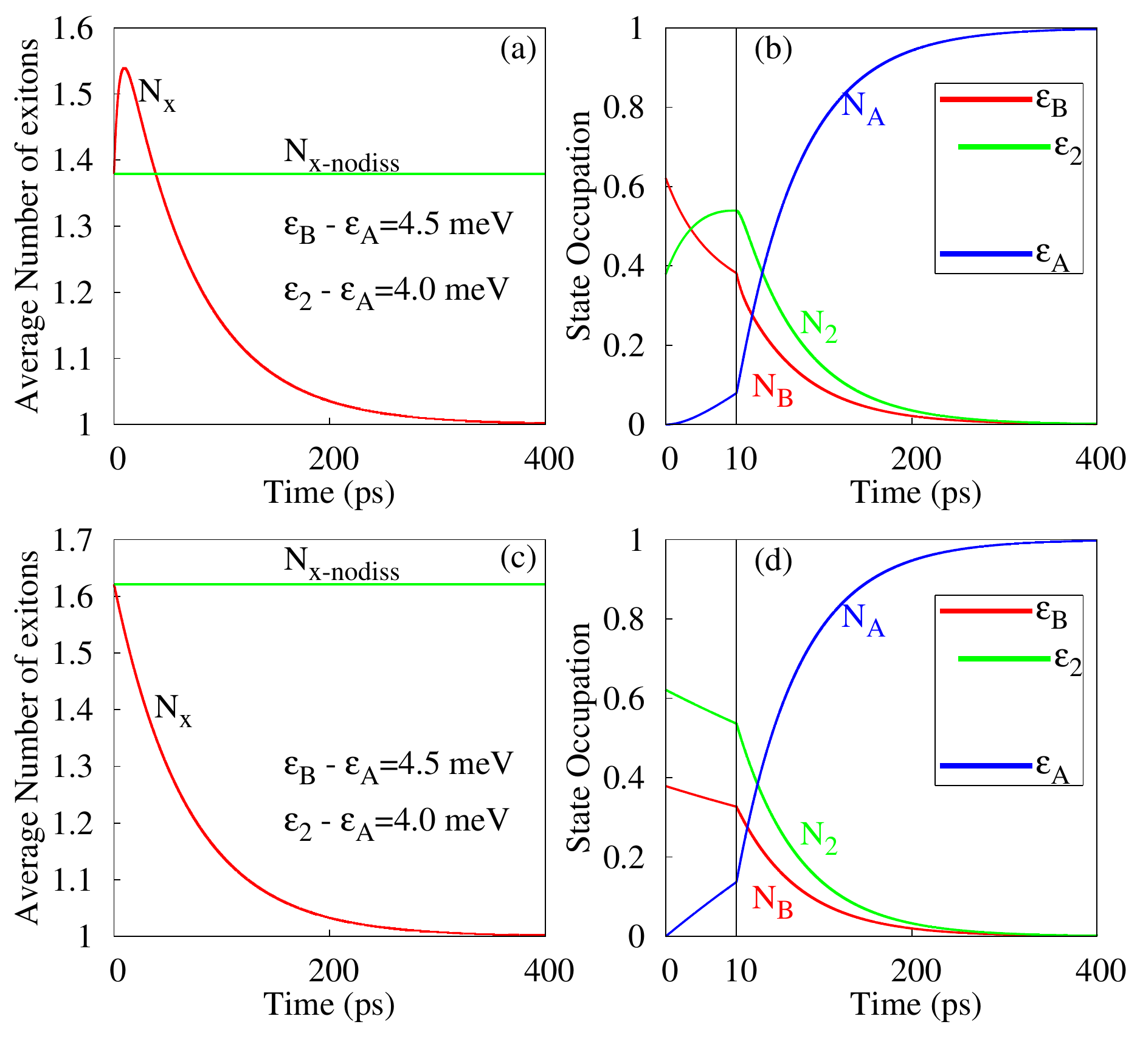}
\caption{\label{Fig5_MA}(Color online)
The variation of state occupation and the average number of excitons 
as a function of time at $T=0$~K. (a,b) for the initial state $|+\rangle$; 
(c,d) for the initial state $|-\rangle$ (narrow band excitation). 
In (b,d), the initial $10$ps of the evolution is expanded. } 
\end{figure}

Under the third possible excitation conditions, when a spectrally 
narrow light field is used, a light beam is tuned to excite just 
the state $|-\rangle$ or $|+\rangle$ (see Appendix for details). 
These two cases are compared in Fig.~\ref{Fig5_MA}. One can see 
that the only difference between starting from $|+\rangle$ or $|-\rangle$  
is the behavior of the BX occupation at the very beginning of the evolution, 
which rises during the first few ps for the initial state $|+\rangle$ 
(Fig.~\ref{Fig5_MA}(b)), which can easily be explained. Since the energy of the state $|+\rangle$ is 
higher than the energy of the state $|-\rangle$, starting with the 
state $|+\rangle$ will cause the transition from $|+\rangle$ to $|-\rangle$. 
After this initial redistribution of occupation, the occupations of the 
states do not depend on the initial conditions. This is in contrast with 
the case without dissipation where the average number of excitons 
remains about 1.4 and 1.6 in the case of starting from $|+\rangle$ and $|-\rangle$, 
respectively, in accordance with the composition of these eigenstates 
in terms of the X and BX states. Obviously, no oscillations can be 
observed in the evolution of the system which was initially prepared 
in one of the eigenstates $|+\rangle$ or $|-\rangle$. 

\begin{figure}[tb]
\centering
\includegraphics[width=85mm]{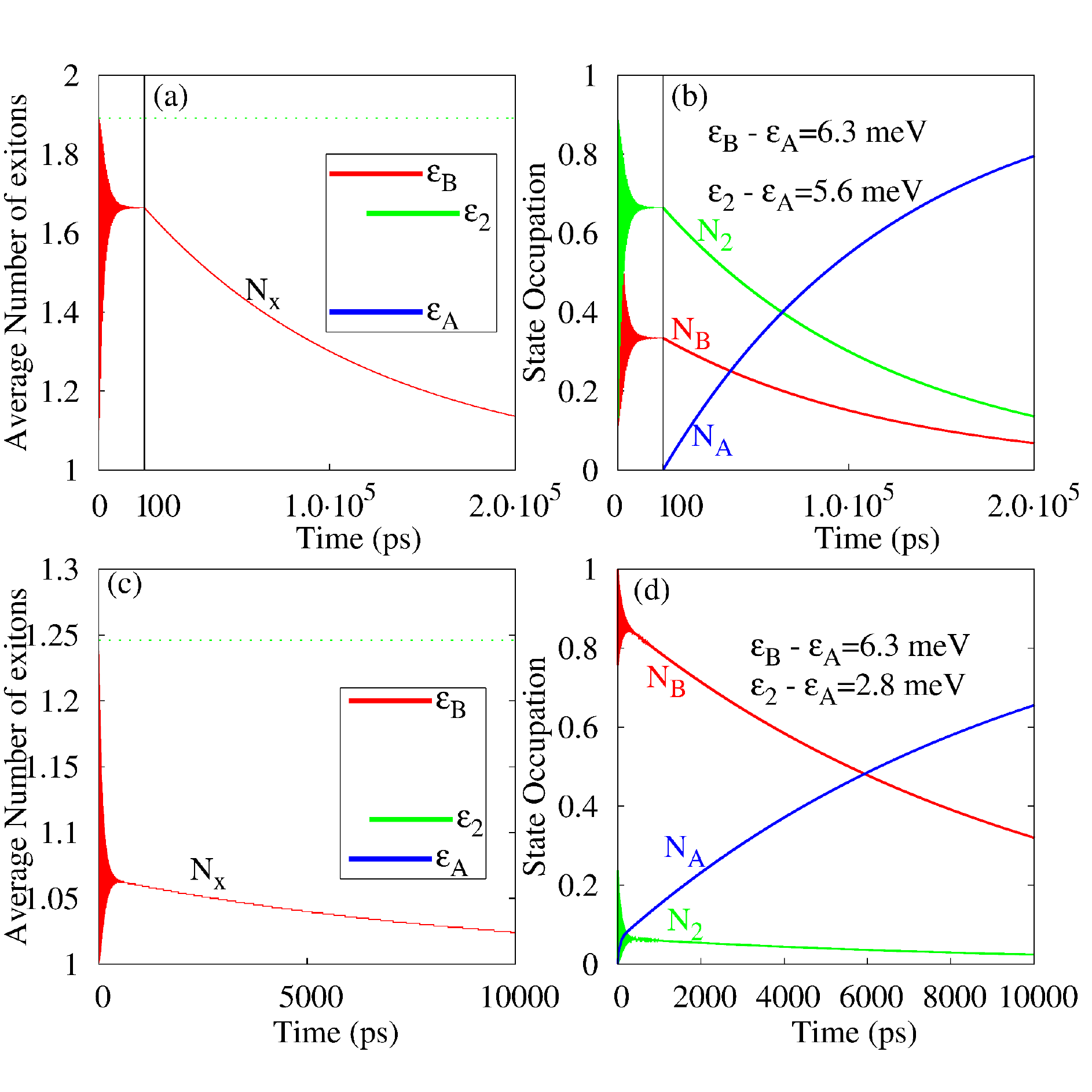}
\caption{\label{Fig6_MA}(Color online)
The state occupations (right panels) and the average number of excitons 
(left panels) for 2 alignments of the energy levels in the case of increased 
energy difference between the states (1.4 times larger than Fig.~\ref{Fig3_MA}) 
at $T=0$~K. In the left panels, the upper envelope of the oscillations of the 
average number of excitons in the case without dissipation is also 
shown (green dotted line). Coherent ultrafast excitation is assumed here. }
\end{figure}

In order to understand the dependence of the observed dynamics on the energy 
spacing between the levels, we will now discuss the case when the energy 
differences between the states are 1.4 times larger than the set of 
parameters in Fig.~\ref{Fig3_MA}. The dissipation parameters $J_0=1~$ps$^{-1}$ 
and $\omega_0=2~$ps$^{-1}$ are kept unchanged. The results of the simulations 
are shown in Fig.~\ref{Fig6_MA}. The impact ionization now appears in a shorter 
time interval (about 10 ps). If the state $|2\rangle$ lies close to the state 
$|B\rangle$ [Fig.~\ref{Fig6_MA}(a,b)], because of the large energy difference 
between the states $|+\rangle$ , $|-\rangle$ and $|A\rangle$ (beyond the cut-off 
energy $\hbar \omega_0$), the rate of the Auger recombination is decreased as 
compared to the previous case and the BX occupation persists much longer. The 
oscillation amplitude is much lower than in the previous case (due to a smaller 
mixing between the X and BX states) and decays rapidly (in a few ps). 
As it is depicted in Fig.~\ref{Fig6_MA}(c,d), when the BX state gets closer to 
the state A, the degree of the impact ionization decreases dramatically. Besides, 
oscillation amplitude decays in a much longer time of a few hundreds of picoseconds. 
The achieved number of excitons is much lower than in all the previously discussed cases.

\begin{figure}[tb]
\centering
\includegraphics[width=85mm]{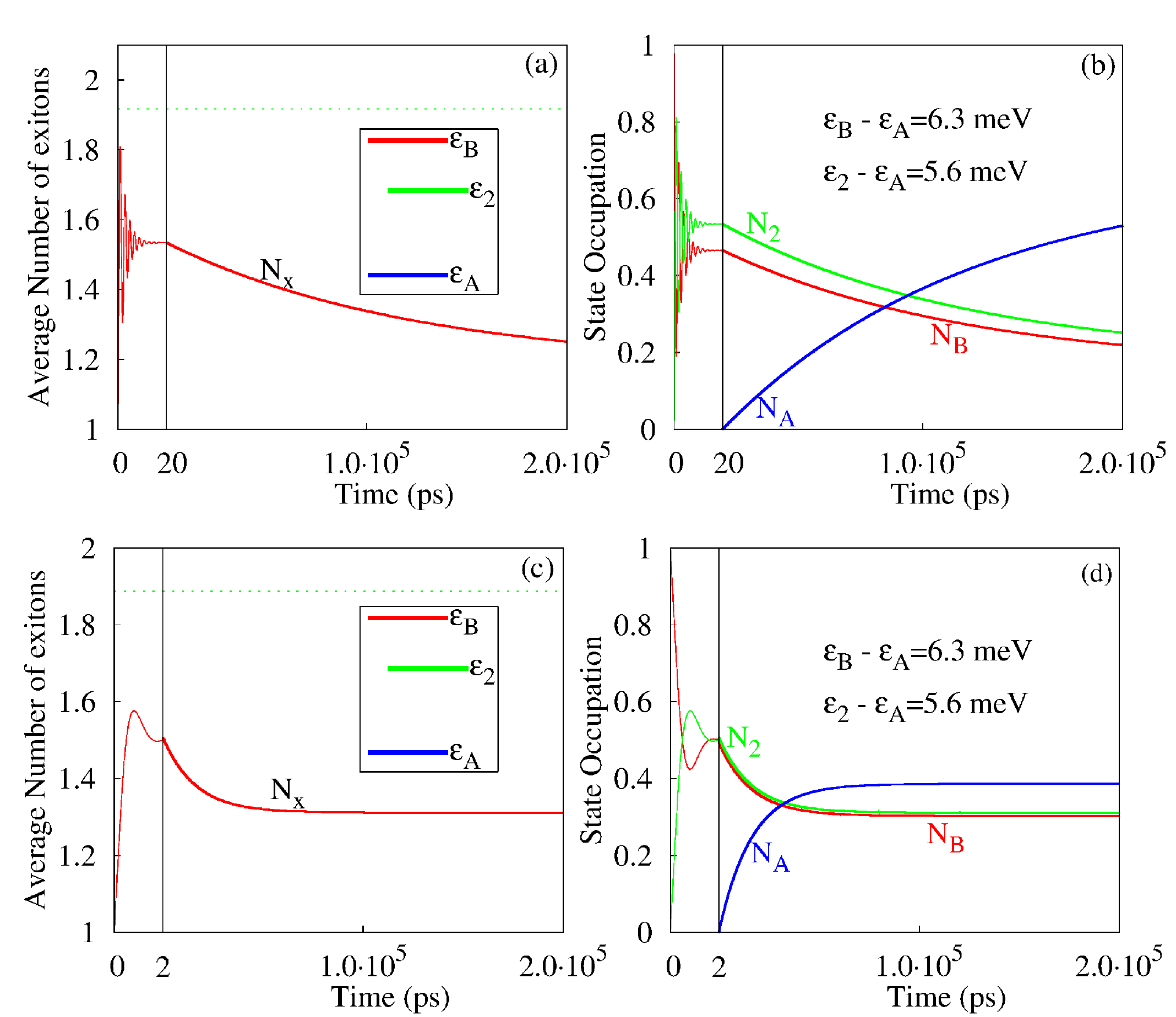}
\caption{\label{Fig650_MA}(Color online)
As in Fig.~\ref{Fig6_MA} but for $T=50$~K (a,b) and $T=300$~K (c,d). } 
\end{figure}

At higher temperature, Fig.~\ref{Fig650_MA}, the impact ionization process takes 
place on the time scale of several picoseconds (decreasing with increasing temperature) 
and is then followed by a slow Auger recombination on the order of nanoseconds 
for this set of parameters).

In a realistic sample, one deals with an ensemble of nanocrystals with a 
certain size dispersion. To obtain the dynamics for the whole nanocrystal 
ensemble, we average our results over the energy of the BX state $|2\rangle$ 
which can be different for each single nanocrystal. We use a Gaussian 
distribution for the value of $\epsilon_2$, 
\begin{eqnarray}
f(\epsilon_2)=\frac{1}{\sqrt{2\pi}\sigma}e^{-\frac{1}{2}\left(\frac{\Delta E}{\sigma}\right)^2},
\end{eqnarray}
where $\Delta E=\epsilon_2-\epsilon_{20}$ and $\sigma$ is a 
standard deviation, while keeping $\epsilon_A$ and $\epsilon_B$ constant.

\begin{figure}[tb]
\centering
\includegraphics[width=85mm]{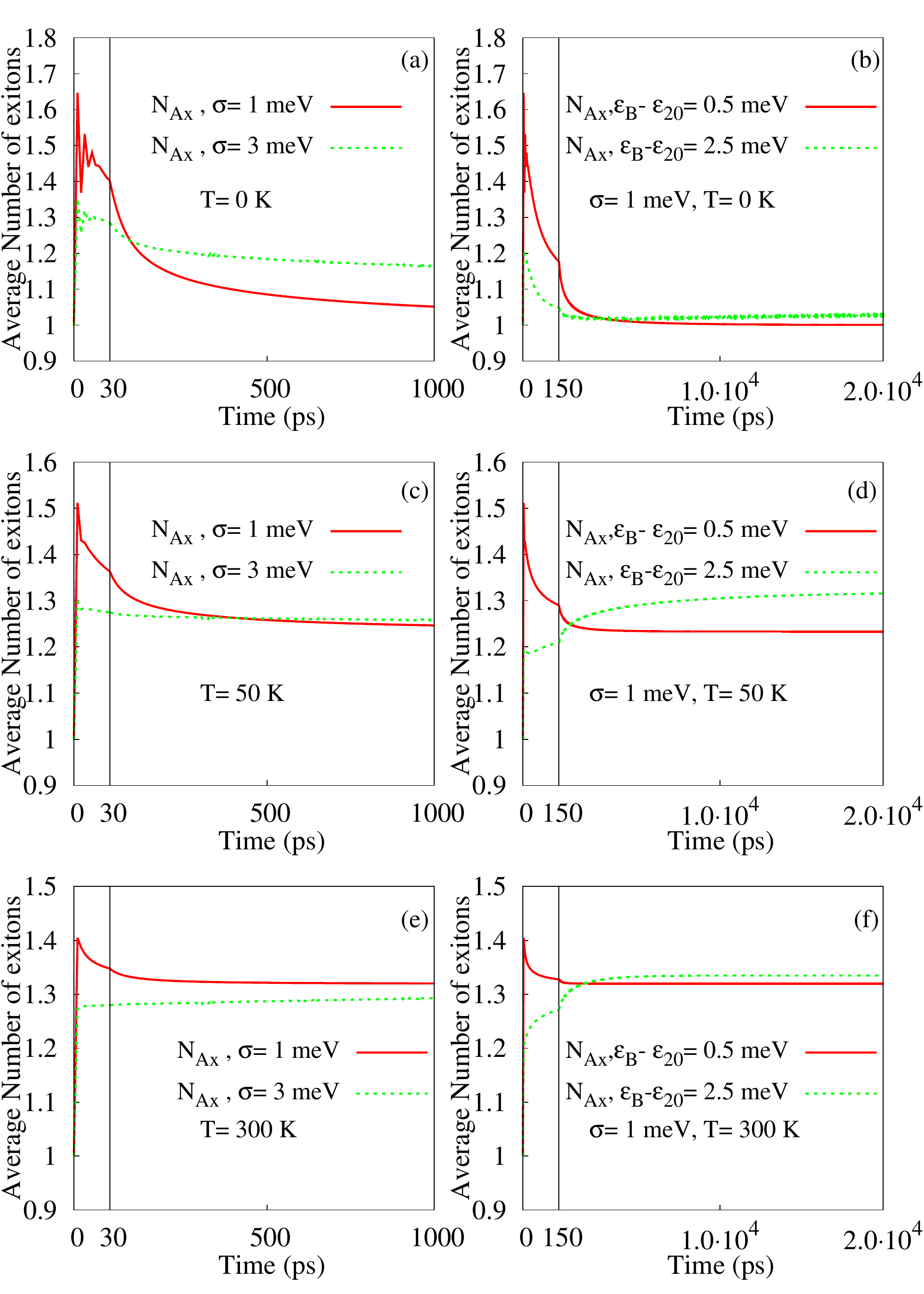}
\caption{\label{Fig7_MA}(Color online)
The variation of the average number of excitons as a function of time at 
three different temperatures. Left panels: for two different values of 
$\sigma$. Right panels: for two different energies of the BX state for 
$\sigma=1$~meV. (a,b) $T=0$~K, (c,d) $T=50$~K and (e,f) $T=300$~K. Ultra 
fast excitation conditions is assumed here.} 
\end{figure}

The results are shown in Fig.~\ref{Fig7_MA} for different sets of parameters. 
In Figs.~\ref{Fig7_MA}(a,c,e), we fix the mean BX energy $\epsilon_{20}$ at 
4 meV and the energy $\epsilon_{B}$ at 4.5 meV above $\epsilon_A$ and show 
the results for three different values of $\sigma$. In Figs.~\ref{Fig7_MA}(b,d,f), 
a small standard deviation, $\sigma=1$~meV is chosen and the results for 
two different level alignments are shown.
    
Comparison of Fig.~\ref{Fig7_MA} with Figs.~\ref{Fig3_MA} and \ref{Fig350_MA}
shows that a small inhomogeneity of the level alignment in the
ensemble (red lines in Figs.~\ref{Fig7_MA}(a,c,e)) does not change the
overall system dynamics. The amplitude of the oscillations at $T=0$ is
reduced due to ensemble dephasing but the average trend is almost
exactly the same. This is true for all the level alignments studied here,
and at all temperatures, as can be seen in
Figs.~\ref{Fig7_MA}(b,d,f). On the other hand, increasing the
inhomogeneity reduces the amplitude of the initial peak of the average
number of excitons, in particular at higher temperatures, and leads to
a nearly featureless time dependence 
of this quantity after the initial ultrafast growth. For larger
inhomogeneities, also the overall (long time) value of the average
number of excitons is reduced. This suppression of the MEG efficiency
in the ensemble in our model is due to the fact that in a strongly
inhomogeneous ensemble the contribution from NCs with very distant
levels becomes larger.
\subsection{Non-Markovian corrections}
\label{sec:non-Markov}

In this section, we discuss some further technical aspects of the modeling of
the MEG dynamics in NCs. We assess the corrections due to the
reservoir memory (beyond the Markov approximation) and study the
differences between Ohmic and super-Ohmic reservoir models. 

\begin{figure}[tb]
\includegraphics[width=85mm]{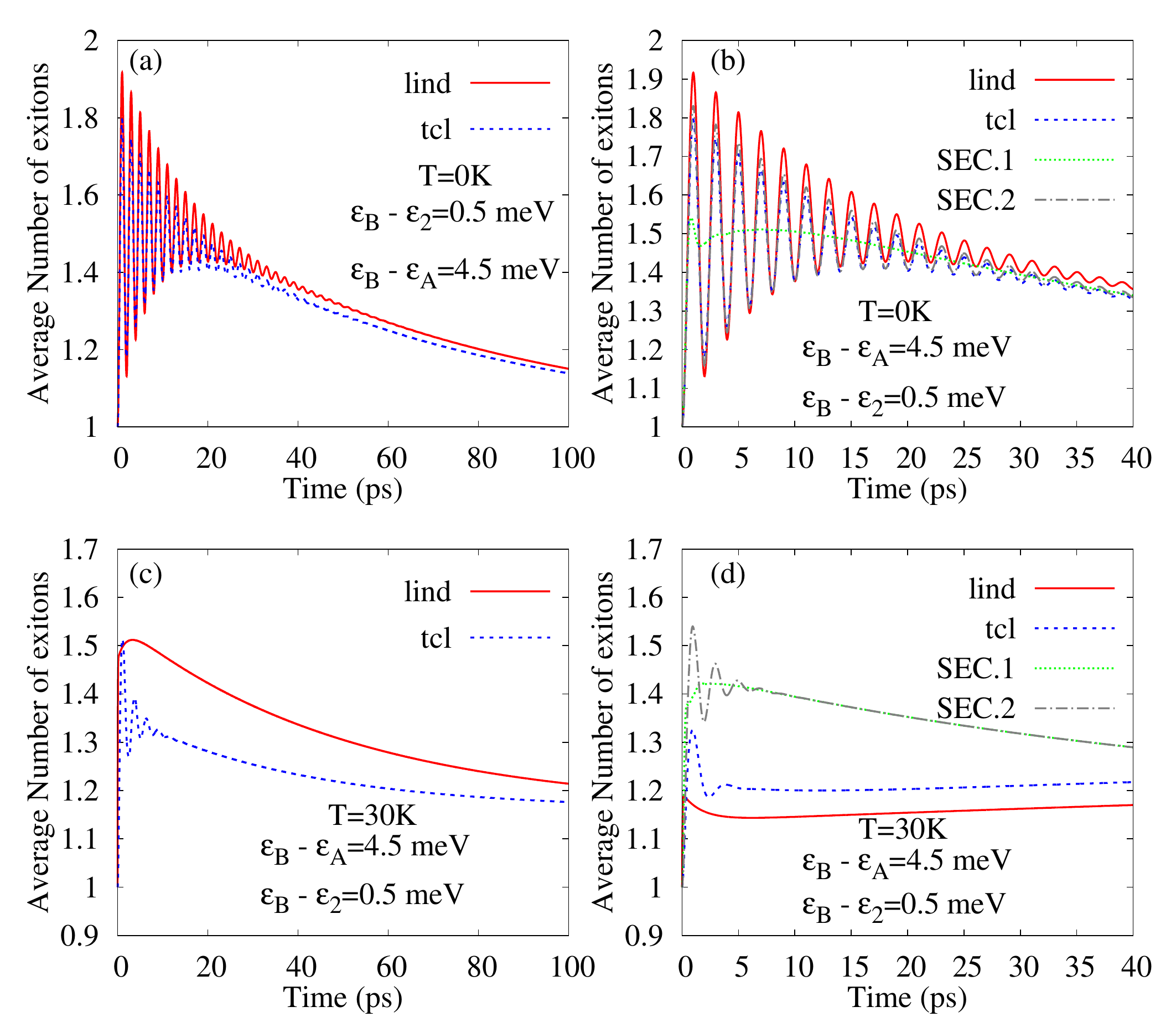}
\caption{Comparison of the simulation results using different
  Markovian and non-Markovian approximations.} 
\label{fig:non-Markov}
\end{figure}

In Fig.~\ref{fig:non-Markov}, we compare the simulation results for
one selected set of system parameters obtained from the Lindblad
equation (red solid lines) and from the non-Markovian TCL equation
(blue dashed lines). In order to understand the role of various
approximations made on the way to the Markovian description in terms
of the Lindblad equation, we present also results
following from two kinds of secular approximations to the TCL
equation (green dotted and gray dash-dotted lines). In the
approximation labeled ``SEC.1'', only the rates $\Gamma_{ijji}(t)$
appearing in the Lindblad equation are kept (but no Markovian
approximation is made, which is reflected on the level of the TCL
equation by the time dependence of the rates). In the second secular 
approximation, denoted ``SEC.2'', also the other subset of secular
rates, $\Gamma_{iijj}(t)$ is retained. At long times, these rates
become proportional to the spectral density at zero frequency, hence
they tend to zero in the zero temperature limit. Therefore, at low temperatures,
they are only important in the initial period of the dynamics (for
times on the order of $1/\omega_{0}$), when non-Markovian effects are
dominant. 

As one can see in Fig.~\ref{fig:non-Markov}(a), at zero temperature
the corrections to the Markovian dynamics are rather small and
essentially amount to a reduced amplitude of the oscillations observed
during the first few tens of picoseconds after the excitation. Both
the time-averaged value of this oscillating exciton number as well as
the long-time asymptotics are nearly the same in both cases. This is
also visible in Fig.~\ref{fig:non-Markov}(b), where the initial period
of time is shown in more detail. A close look at the curves reveals
that the difference between the Lindblad and TCL results is due to
initial damping within a few picoseconds from the initial time, while
the subsequent evolution is characterized by the same exponential
damping of the oscillations and decay of the populations in both
cases. This initial damping is due to the fact that the short-time values
of the non-Markovian damping rates involve interaction with the whole reservoir
spectrum, while in the Markov limit only the resonant modes are
involved, corresponding to the relatively low values of the spectral
density in its high-energy tails in the present case.

In Fig.~\ref{fig:non-Markov}(b), we have also displayed the results
obtained from the non-Markovian TCL equations in which the
non-secular (oscillating) terms have been removed. It is clear that both
classes of secular terms must be kept in the non-Markovian
description even though only one of them appears in the Lindblad
equation which yields quite accurate results. Otherwise, the initial,
non-Markovian damping is overestimated by an order of
magnitude. Still, however, the trend is reproduced correctly. Similar
effect, although with a larger quantitative difference in the
oscillation amplitudes, is seen in a system with lower energy spacing
between the levels (not shown here).

A similar comparison for $T=30$~K, presented in
Fig.~\ref{fig:non-Markov}(c), shows a larger difference between the
Markovian and non-Markovian modeling results. Now, the Markov
approximation overestimates the damping of the initial oscillations
and yields a higher number of excitons (the latter depends on the
level alignment; we have observed an opposite situation for smaller
inter-level spacings, not shown here). Much more interesting is the
analysis of the role of secular vs. Markov approximations in this
moderate temperature case, shown in
Fig.~\ref{fig:non-Markov}(d). Here, both secular approximations yield
the same long-time asymptotics but differ in the way the initial
oscillations are reproduced: if only the terms entering in the
Lindblad equation are kept the oscillations vanish completely, as is
also the case for the Lindblad description. The other, more complete secular
approximation, reproduces the oscillations qualitatively correctly,
although with a shift of the average trend. Even in this case,
however, the quantitative characteristics of the MEG process are very
similar in all the approximations. 

\begin{figure}[tb]
\includegraphics[width=85mm]{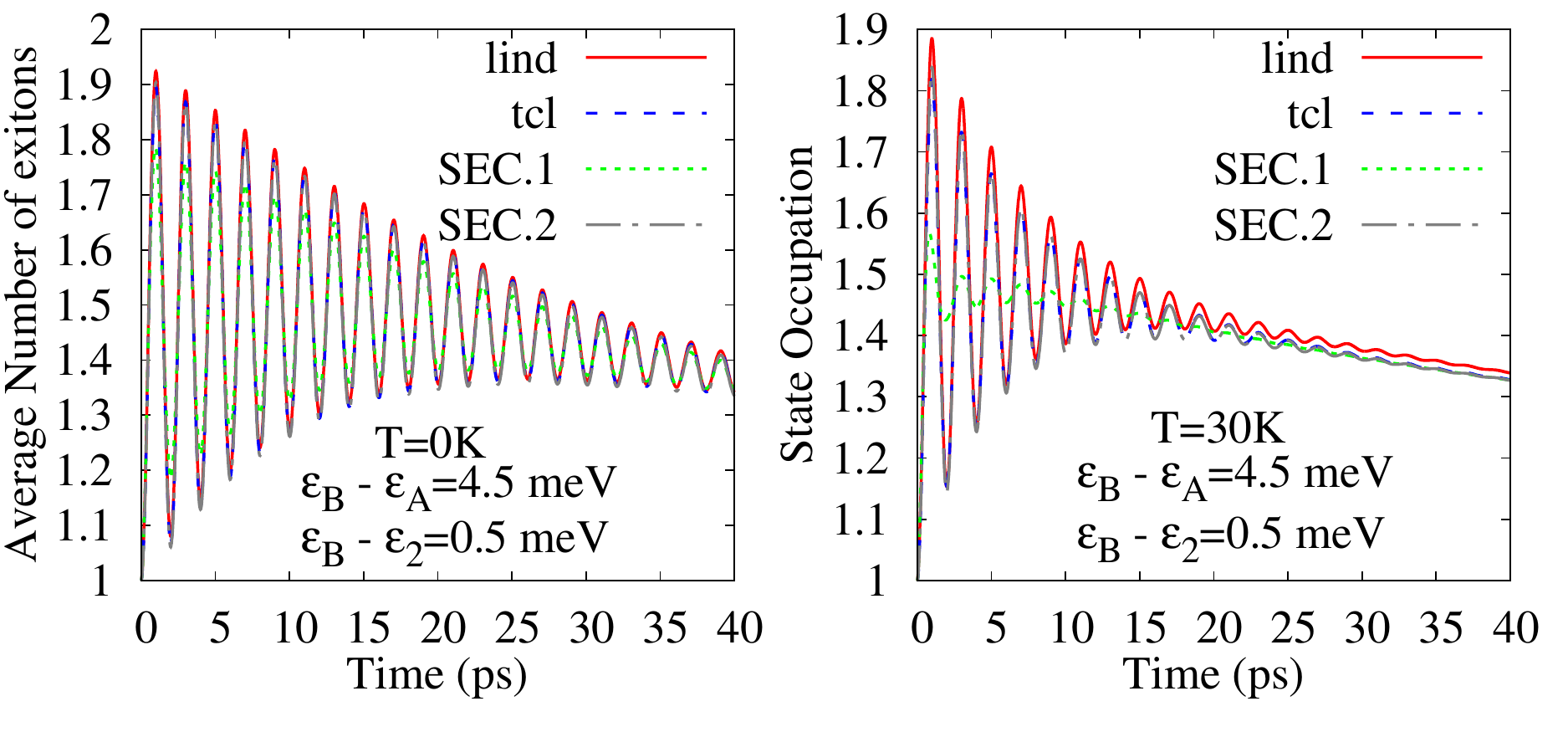}
\caption{Comparison of the   Markovian and non-Markovian
  approximations for an Ohmic and super-Ohmic reservoir.} 
\label{fig:ohmic}
\end{figure}

So far, in all our simulations we have assumed a reservoir with Ohmic
spectral characteristics ($J(\omega)\sim\omega$ at $\omega\to
0$). This is the standard choice for generic modeling of the dynamics
of an open quantum system in the absence of any detailed knowledge
about the particular reservoir in question and has also been employed
in some studies of the dissipative MEG process
\cite{velizhanin11a}. However, some specific reservoirs are known to
have other spectral characteristics. In particular, the
three-dimensional acoustic phonon reservoir shows a super-Ohmic
behavior with $J(\omega)\sim\omega^{3}$ at $\omega\to 0$. This leads
to vanishing spectral densities $R(\omega)$ at $\omega=0$ and, as
mentioned above, to suppression of a class of damping rates in the long
time limit. It seems interesting to study how this qualitative
difference affects the system dynamics both in the Markov limit and in
the non-Markovian model. To this end, in Fig.~\ref{fig:ohmic}, we
show the simulation results for the same system as in
Fig.~\ref{fig:non-Markov} but with a super-Ohmic spectral density 
$J(\omega)=J_{0}(\omega/\omega_{0})^{3}exp(-\omega^{2}/\omega_{0}^{2})$
with the amplitude $J_{0}$ and cut-off frequency $\omega_{0}$ chosen
in such a way that the Markovian transition rates from the states
$|+\rangle$ and $|-\rangle$ to the state $|A\rangle$ are the same as
in the previous Ohmic case. The zero temperature results presented in
Fig.~\ref{fig:ohmic}(a) show a very similar behavior to the Ohmic case
at the same temperature [Fig.~\ref{fig:non-Markov}(b)] but much less
sensitivity to the approximation used. In particular comparison
between the results obtained with various secular approximations shows
that the dynamics is determined by the Lindblad rates
$\Gamma_{ijji}$. This can be attributed to the reduced magnitude of
the spectral density in the vicinity of $\omega=0$, which reduces the
role of the other secular rates not only in the long time limit but
also already for shorter times. This role of the secular terms of the
second type is larger at higher temperatures
[Fig.~\ref{fig:ohmic}(b)], where neglecting them leads to strongly
overestimated damping. Remarkably, the non-Markovian corrections are
much smaller in the super-Ohmic case and both the average trend of the
evolution of the exciton number as well as the long-time value are
nearly the same here, in contrast to the Ohmic model (especially at
higher temperatures).

\section{Conclusions}
\label{sec:concl}

We have formulated and studied a few-level model of dissipative
multiple exciton generation and relaxation dynamics following a photon
absorption in a semiconductor nanocrystal. We have accounted for the
interaction with the reservoir by introducing a physically
motivated set of spectral densities which allowed us to relate the
impact ionization and Auger relaxation rates to the diagonal and
off-diagonal carrier-reservoir couplings, respectively, and to
highlight the role of the interband Coulomb coupling for the magnitudes
of the rates governing these two processes. 

With this model, we have solved the Markovian quantum Master (Lindblad)
equation to investigate the impact of dissipation on the evolution of
the single- and bi-exciton occupation. We have shown
that the system evolution strongly changes if the dissipation is
included. In many cases, the maximum average number of electron-hole pairs
(i.e., the efficiency of the MEG process) is increased if dissipative
transitions are allowed and can be close to 2.  
Thus, dissipation can play a constructive role in the MEG process and
is not neccessarily a competing factor
\cite{shabaev06,witzel10,allan06}. 
In a
certain range of parameters, the growth of the exciton number (MEG
process) is very fast (on picosecond time scale) and the following
decay of the biexciton population
(Auger process) is much slower (on the time scale of hundreds of
picoseconds), which means that such a dissipative dynamics following
an ultrafast excitation cannot be excluded based on the observed very
fast time scale of the process \cite{schaller05}. In addition, the
differences between 
the system dynamics under various excitation conditions, which are
present in the dynamics of an isolated nanocrystal, are washed out by
fast reservoir-induced dephasing. We have verified also that similar
dissipation-related features in the system kinetics are observed in an
inhomogeneous ensemble of nanocrystals.

We have studied also the sensitivity of the modeling results to
various formal characteristics of the model. We have shown that the
dynamics very strongly depends on the position of the high-frequency
cut-off of the reservoir spectral density. The simulated dynamics
depends to some extent on the chosen class of the reservoir models but
the observed differences between the Ohmic and super-Ohmic models are
mostly of quantitative character and change neither the qualitative
features of the dynamics nor the quantitative expectations for an
overall MEG yield in a nanocrystal ensemble. We have investigated also
the role of non-Markovian corrections to the system dynamics. Although
the evolution found from non-Markovian equations differs in some
cases from that obtained in the Markov limit these discrepancies
mostly have the form of oscillations that are present only during the
first few tens of picoseconds after excitation and are not expected to
affect the overall quantitative predictions for the MEG yield.

\section{Acknowledgments}
This work was supported by the TEAM programme of the Foundation 
for Polish Science co-financed from the European Regional Development Fund.

\section{Appendix: The density matrix elements corresponding to various excitation conditions}

In this appendix, we derive the state occupations after optical 
excitation in various excitation conditions. First, we consider 
the interaction between a strong coherent field (as a pulsed 
excitation treated classically) and our 4 levels system. 
A single mode radiation source, such a laser, will produce an 
electromagnetic wave with amplitude $\mathcal{E}_0(t)$ and frequency $\Omega$,
\begin{eqnarray}
\boldsymbol{\mathcal{E}}(t)=\boldsymbol{\mathcal{E}_0}(t) \cos \Omega t.
\end{eqnarray}
The interaction Hamiltonian between this electromagnetic wave and our system is
\begin{eqnarray}
H_{\mathrm{int}}=-\boldsymbol{d}\!\ \cdot\!\  \boldsymbol{\mathcal{E}}(t)=f(t) \cos \Omega t (|G\rangle\!\langle B|+ \mathrm{h.c.}),
\end{eqnarray}
where we defined $f(t)=-\boldsymbol{d}\!\ \cdot\!\  \boldsymbol{\mathcal{E}_0}(t)$. 
In the eigenbasis $\left(|+\rangle, |-\rangle\right)$ and in the rotating wave approximation, we have
\begin{eqnarray}
H_{\mathrm{int}}=\frac{1}{2}f(t)(|G\rangle\!\langle +|e^{i\Delta_+t}\!\cos\frac{\theta}{2}\nonumber\\-|G\rangle\! \langle -|e^{i\Delta_-t}\!\sin\frac{\theta}{2}+ \mathrm{h.c.}),
\end{eqnarray}
where $\Delta_{\pm}=\Omega-\frac{E_{\pm}}{\hbar}$ is the detuning 
from the transition energy. 
The system state after the optical pulse up to the $2$nd order in $\mathcal{E}_0(t)$ is
\begin{eqnarray}
\rho&=&\rho_0 -\frac{i}{\hbar} \int_{\!-\infty}^\infty dt \left[H_{\mathrm{int}},\rho_0\right]\nonumber\\
&&-\frac{1}{2\hbar^2} \int_{\!-\infty}^\infty dt\int_{\!-\infty}^t d\tau  \left[H_{\mathrm{int}}\left[H_{\mathrm{int}},\rho_0\right]\right],
\end{eqnarray}
where $\rho_0=|G\rangle\!\langle G|$.
The occupations of the $|+\rangle\ $and $|-\rangle\ $states appear in 
the $2$nd order term, assuming a Gaussian envelope for $f(t)$,
\begin{eqnarray}
 f(t)=\frac{1}{\sqrt{2\pi}\tau}e^{-\frac{1}{2}\left(\frac{t}{\tau}\right)^2},
\end{eqnarray}
one finds the density matrix elements corresponding to the coherent excitation in the form
\begin{eqnarray}\label{coherent}
\langle\pm|\rho|\pm\rangle&=&\frac{1}{\sqrt{2\pi}\tau}\frac{1{\pm}\cos\theta}{2} e^{-\tau^2\Delta_{\pm}^2}\nonumber\\
\langle+|\rho|-\rangle&=&\langle-|\rho|+\rangle=-\frac{1}{2\sqrt{2\pi}\tau}\sin\theta e^{-\frac{\tau^2}{2}\left(\Delta_+^2+\Delta_-^2\right)}
\end{eqnarray}
where $\tau$ is the pulse duration.

In the case of a narrow band excitation condition, when $\Delta_{\pm}  \tau \gg 1$, 
all the exponents in Eq.~\eqref{coherent} vanish except for the one corresponding 
to $\Delta_{\pm}=0$. Thus the only non-zero element will be $\langle-|\rho|-\rangle$ 
or $\langle+|\rho|+\rangle$ corresponding to the laser tuned to $\Delta_-$ or $\Delta_+$, respectively. 
On the other hand, for a broad band excitation condition (short pulse), 
$\Delta_{\pm}  \tau \ll 1$, all the exponents in Eq.~\eqref{coherent} are almost 
equal to 1. After inverting Eq.~\eqref{pm} and substituting to Eq.~\eqref{coherent} 
one finds $\langle B|\rho|B\rangle=1$ and $\langle2|\rho|2\rangle=0$. Hence, 
under this conditions, only the state $|B\rangle$ is excited.

Second, let us consider broad band thermal radiation. Then an incoherent mixture 
of system eigenstates is excited, so that there is no coherence between the $|+\rangle\ $
and $|-\rangle\ $ states. The interaction Hamiltonian is described by 
\begin{eqnarray}
\lefteqn{H_{\mathrm{int}}=\boldsymbol{d}\!\ \cdot\!\  \boldsymbol{\mathcal{E}}=‪\sum_{k\lambda}g_{k\lambda}(b_{k\lambda}+b_{k\lambda}^{\dag})}\\
&&\times(|G\rangle\!\langle A|+\!\cos\left(\frac{\theta}{2}\right)|G\rangle\!\langle +|-\!\sin\left(\frac{\theta}{2}\right)|G\rangle\!\langle -|+\mathrm{h.c.}),\nonumber
\end{eqnarray}
where $b_{k\lambda}$ and $b_{k\lambda}^{\dag}$ are photon annihilation 
and creation operators respectively. 
The occupation of the system states resulting from this kind of excitation 
are proportional to the corresponding transition states, which can be found 
using the Fermi golden rule. Since the states $|+\rangle\ $and $|-\rangle\ $ 
are very close compared to the photon energy, the difference in the photon 
density of states and coupling magnitude is negligible and one finds up to a constant.
\begin{eqnarray}
\langle\pm|\rho|\pm\rangle &\sim& \frac{1{\pm}\cos\theta}{2}, \nonumber\\
\langle+|\rho|-\rangle&=&\langle-|\rho|+\rangle=0.
\end{eqnarray}

\bibliographystyle{prsty}
\bibliography{abbr,quantum2}

\end{document}